\documentclass[%
 aip,
 amsmath,amssymb,
 preprint,%
]{revtex4-1}

\usepackage{graphicx}
\usepackage{dcolumn}
\usepackage{bm}

\usepackage[utf8]{inputenc}
\usepackage[T1]{fontenc}
\usepackage{mathptmx}
\usepackage{subfigure}
\usepackage{hyperref}

\begin{document}

\title{Thickness Dependence of Magneto-transport Properties in Tungsten Ditelluride} 



\author{Xurui Zhang}
\email{zxx150430@utdallas.edu}
\affiliation{Department of Physics, The University of Texas at Dallas, Richardson, TX 75080, USA}

\author{Vivek Kakani}
\affiliation{Department of Physics, The University of Texas at Dallas, Richardson, TX 75080, USA}

\author{John M. Woods}
\affiliation{Department of Mechanical Engineering and Materials Science, Yale University, New Haven, CT 06511, USA}

\author{Judy J. Cha}
\affiliation{Department of Mechanical Engineering and Materials Science, Yale University, New Haven, CT 06511, USA}

\author{Xiaoyan Shi}
\email{xshi@utdallas.edu}
\affiliation{Department of Physics, The University of Texas at Dallas, Richardson, TX 75080, USA}


\begin{abstract}

We investigate the electronic structure of tungsten ditelluride (WTe$_2$) flakes with different thicknesses in magneto-transport studies. The temperature-dependent resistance and magnetoresistance (MR) measurements both confirm the breaking of carrier balance induced by thickness reduction, which suppresses the `turn-on' behavior and large positive MR. The Shubnikov-de-Haas oscillation studies further confirm the thickness-dependent change of electronic structure of WTe$_2$ and reveal a possible  temperature-sensitive electronic structure change. Finally, we report the thickness-dependent anisotropy of Fermi surface, which reveals that multi-layer WTe$_2$ is an electronic 3D material and the anisotropy decreases as thickness decreases.

\end{abstract}


\maketitle 


\section{Introduction}

Tungsten ditelluride (WTe$_2$), a layered transition metal dichalcogenide (TMD) material, has attracted a great deal of interests due to its unique electronic transport properties since the discovery of the non-saturating positive large magnetoresistance (MR) in bulk \cite{ali2014large}. It is widely believed that the extraordinary MR comes from the nearly perfect balance between electron and hole concentrations \cite{ali2014large, alekseev2015magnetoresistance, pletikosic2014electronic, thoutam2015temperature, wu2015temperature, dai2015ultrafast, luo2015hall}. Many other peculiar electronic properties have also been observed in WTe$_2$ in transport measurements, such as `turn-on' behavior \cite{ali2014large, wang2015origin, woods2017suppression, ali2015correlation}, multi-Fermi pockets revealed by Shubnikov-de-Haas (SdH) oscillations \cite{cai2015drastic, zhu2015quantum, xiang2018thickness}, surprisingly small Fermi surface anisotropy \cite{thoutam2015temperature}, ferroelectric property \cite{fei2018ferroelectric, sharma2019room, liu2019vertical, yang2018origin}, superconductivity \cite{fatemi2018electrically, pan2015pressure, kononov2021superconductivity}, etc. WTe$_2$ is also a topological material. In the bulk form, WTe$_2$ has been predicted \cite{soluyanov2015type} and observed \cite{li2017evidence} to be a type-II Weyl semimetal. In the monolayer form, WTe$_2$ is a quantum spin Hall insulator \cite{wu2018observation, qian2014quantum} at low carrier density (n). However, it is still unclear how the topological property and the electronic structure evolve by reducing the crystal thickness.

Here we investigate the evolution of electronic structure in WTe$_2$ flakes with different thicknesses by performing temperature dependent resistance ($R$-$T$) measurements, MR measurements, SdH oscillation studies and angle-dependent MR measurements. Experiments show that the imbalance of carrier densities caused by thickness reduction plays an important role in the suppression of the `turn-on' behavior and the large positive MR, while the non-saturating characteristic was hardly affected. We further confirm that the multi-layer WTe$_2$ is also an electronic 3D material like bulk crystal and the anisotropy reduces as thickness decreases.

\section{Experiment}

The WTe$_2$ flakes with different thicknesses were obtained by mechanical exfoliation of bulk WTe$_2$ crystals synthesized by chemical vapor transport \cite{woods2017suppression}. Six exfoliated flakes characterized in this paper can be classified into three groups. The first group contains two thick samples (denoted as `sample 1' and `sample 2' hereafter) with thickness around 150~nm, the second group contains three thin flakes with thickness around 20~nm (denoted as `sample 3' through `sample 5' hereafter), and the third group contains one ultra-thin flake with thickness at 5~nm (`sample 6'), whose transport properties have been reported elsewhere \cite{zhang2020crossover}. In this paper, we will focus on the first two groups. The thick flakes are directly transferred onto silicon substrates with 285~nm-thick SiO$_2$ coating on surface. The thin flakes was encapsulated between two pieces of hexagonal boron nitride (hBN) thin flakes which were about 10~nm thick and transferred onto the substrates by dry transfer technique \cite{castellanos2014deterministic}. Thin WTe$_2$ flakes get oxidized easily upon exposure to air \cite{Fatemi926, wang2015tuning}. Hence the hBN flakes are necessary here to protect the thin samples from air-induced degradation. In addition, hBN flakes provide a cleaner interface for WTe$_2$. For the thick samples, photo-lithography was used to make the patterns. For the thin samples, electron-beam lithography was used to make patterns. The Ohmic contacts were deposited by electron-beam evaporation of Pd/Au (10 nm/200 nm for thick samples, 10 nm/40 nm for thin samples) followed by a lift-off process. Transport measurements down to 0.02 K were carried out in an Oxford dilution refrigerator. Both the longitudinal resistance $R_{xx}$ and Hall resistance $R_{xy}$ were measured simultaneously by using standard low frequency lock-in techniques.

\section{$R$-$T$ characteristics and `Turn-on' behavior}

Except in the ultra-thin sample, the temperature-dependent longitudinal resistances ($R$-$T$) in all other five devices (samples 1-5) with different thicknesses show metallic properties across the full experiment temperature range, as shown in FIG. 1(a). The ultra-thin sample (sample 6), however, shows an insulating temperature dependence at low temperature region and it is more than 10 times resistive than all other samples. All samples in the same group show similar behavior. Thus, we pick up sample 2 and sample 3 as representatives for the thick and thin groups, respectively. In FIG. 1(b), the $R$-$T$ curves for sample 2 and sample 3 are plotted in log-log scale and Fermi liquid fits, $R_0=\alpha+\beta T^2$, applied at low temperatures are also shown as the black dashed lines for both samples. Here $R_0$ represents the resistance at 0 T magnetic field and $\alpha, \beta$ are two fitting parameters. We found that the maximum applicable $T$ of the Fermi liquid fit decreased from $\sim 60$ K in sample 2 to $\sim 40$ K in sample 3 with decreasing thickness. Comparing with the case in bulk ($\sim 80$ K) \cite{wang2015origin}, the trend shows consistency that the applicable $T$ range decreases with thickness decreasing.

We further investigated the $R$-$T$ curves at various magnetic fields up to 12 T. We unambiguously observed the `turn-on' behavior in sample 2, as shown in FIG. 1(c). The $R$-$T$ curves gradually change from metallic to insulating with magnetic field increasing from 0 T to 12 T. The critical field $\mu_0 H^*$ is around 7 T, which is much larger than the one reported in bulk (below 2 T) \cite{ali2014large, wang2015origin, woods2017suppression}. Due to the fact that the `turn-on' behavior only occurs in the Fermi liquid state \cite{wang2015origin, khveshchenko2001magnetic}, we assume that the larger $H^*$ in our case is caused by the shift of the Fermi liquid state to a lower temperature region due to the thickness decreasing. Our assumption can be supported by a comparison measurement in sample 3. In sample 3, the `turn-on' behavior can't be observed up to 12 T. In a higher field, SdH oscillations emerge and disguise the `turn-on' behavior. That means a magnetic field larger than 12 T is required to manifest the `turn-on' behavior with the Fermi liquid state further moving to a lower temperature region (below 40 K).

In order to investigate the origin of the `turn-on' behavior, we applied Kohler’s rule:
\begin{equation}\label{eq:one}
\rm MR=\it A(H/R_0)^m
\end{equation}
in which $R_0$ is the resistance of $R$-$T$ curve at zero magnetic field and $A, m$ are the constant fitting parameters. Here MR is defined as MR=$[R(T, H)-R_0 (T)]/R_0(T)$. We found that the R-T curves in FIG. 1(c) can be scaled into one curve. Specifically, the Kohler’s rule fitting gave the parameter $m=1.8$. Comparing with the case in bulk ($m=1.92$) \cite{wang2015origin}, $m$ decreased and deviated further from the perfect carrier compensation of $m=2$ \hfill\cite{sondheimer1947theory, chan2014plane, murzin1998effect}. The scaling behavior obtained by Kohler’s rule confirmed that the $R$-$T$ curves at different magnetic fields have the same temperature dependence, although it looks like the curve at higher field shows larger MR effect. This is quite different from the case of a magnetic field-induced metal-insulator transition, which requires a larger increasing rate at a higher magnetic field due to gap opening \cite{khveshchenko2001magnetic, kopelevich2006universal}. In addition, our results show that the `turn-on' behavior takes its origin from carrier compensation. The nearly broken carrier compensation in sample 3 ($m=1.69$) makes the `turn-on' behavior almost invisible even at 12 T.

\section{Two-band model fittings}

To further confirm whether perfect carrier compensation is related to the `turn-on' behavior or not, we performed MR measurements on both samples. The MR curves of both samples at different temperatures are shown in FIG. 2(a) and 2(d). We found that the MR curves in sample 2 crossed at two points located at 7 T and -7 T, respectively. This is consistent with the $\mu_0 H^*$ of the `turn-on' behavior. However, it's almost impossible to identify such two crossing points in sample 3 since the MR curves at different temperatures are fully overlapped in large magnetic field region. This is also consistent with the fact that we didn't observe obvious `turn-on' behavior in sample 3.

Quasi-quadratic positive MR can be seen in both samples, which has been attributed to the perfectly balanced electron and hole densities \cite{pletikosic2014electronic, woods2017suppression, thoutam2015temperature, wu2015temperature, dai2015ultrafast, luo2015hall}. Comparing with the results in bulk \cite{ali2014large}, which recorded MR as high as 13,000,000 $\%$, the highest MR in our samples only showed 1,200 $\%$ due to reduction in thickness. The MR in both samples is notably suppressed, which indicates an imperfect balance between carrier densities and more impact from defects. But there's still no observed trend towards saturation of MR in the samples. In order to examine the balance between carrier densities further, we additionally performed Hall measurements. Combined with MRs, we could extract the carrier densities and mobilities using the two-band model \cite{ashcroft1976solid, rullier2009hall, rullier2010hole}:
\begin{eqnarray}\label{eq:two}
R_{xx} &=& \frac{\sigma_1+\sigma_2+(\sigma_1\sigma_2^2R_{H_2}^2+\sigma_2\sigma_1^2R_{H_1}^2)B^2}{(\sigma_1+\sigma_2)^2+\sigma_1^2\sigma_2^2(R_{H_1}+R_{H_2})^2B^2} \\
R_{xy} &=& \frac{R_{H_1}\sigma_1^2+R_{H_2}\sigma_2^2+\sigma_1^2\sigma_2^2R_{H_1}R_{H_2}(R_{H_1}+R_{H_2})B^2}{(\sigma_1+\sigma_2)^2+\sigma_1^2\sigma_2^2(R_{H_1}+R_{H_2})^2B^2}
\end{eqnarray}
in which $\sigma_i=n_ie\mu_i\ (i=1, 2)$ are the conductance contributions from electron and hole, respectively. $R_{H_i}=1/n_ie \  (i=1, 2)$ are the Hall coefficients for electron-dominant and hole-dominant Hall effect. $n_i$ and $\mu_i$ are carrier densities and mobilities, respectively. The fitting results and the extracted carrier densities and mobilities are shown in FIG. 2.

The two-band model was perfectly applicable to our MR and Hall measurements for sample 2, while there's a little mismatch in the low field region of MR for sample 3 (Fig. 2(e)). The low-field mismatch might come from the disorder-induced quantum interference effect \cite{zhang2020crossover, lu2011competition, hikami1980spin, lu2014finite}, which is more pronounced in ultra-thin sample where weal localization and weak anti-localization effects are stronger. However, this small mismatch doesn't have decisive influence on the extraction of carrier densities and mobilities. In sample 2, we found that the hole density was almost unchanged with temperature while the electron density gradually increases with temperature increasing (Fig. 2(c)). Such a trend is consistent with other transport and ARPES studies \cite{pletikosic2014electronic, thoutam2015temperature, wu2015temperature, dai2015ultrafast, luo2015hall}. However, the electron and hole densities are not completely equal at low temperature in our samples. The charge carrier density ratio, $n_e/n_h$, is about 1.15 in sample 2 and becomes even larger in sample 3, which is 1.26. It indicates that the perfect balance between electron and hole carrier densities will be gradually broken as the thickness decreases. Furthermore, such a growing charge imbalance will cause the MR to be continuously suppressed and will also cause the `turn-on' behavior to become insignificant and eventually disappear. However, it seems that the MR can still maintain the non-saturating characteristic at large field up to 12 T, no matter how much the MR is suppressed. While the MR in some imbalanced-carrier systems tends to saturate eventually at high magnetic fields \cite{yang1999large, kopelevich2003reentrant}, it is not observed in our experiments. This could be attributed to the fact the either the magnetic field used in our experiment is not large enough or the origin of the non-saturating characteristic is not the perfect balance of electron and hole compensation. Actually a linear non-saturating MR has been observed in some Dirac semimetals \cite{liang2015ultrahigh, liu2014discovery, shekhar2015extremely, ghimire2015magnetotransport, xu2015discovery}, which is believed to take the origin of the lifting of a remarkable protection mechanism induced by time reversal symmetry that strongly suppresses backscatterings at zero magnetic fields. Similar linear MR curves have been observed in disordered WTe$_2$ flakes \cite{zhang2020crossover, liu2016effect}. Possible link between the non-saturating MR in WTe$_2$ and such a mechanism also deserves further investigation.

It is worth mentioning that both the electron and hole mobilities in sample 2 increased with decreasing temperature, which might be due to the suppression of phonon scattering at lower temperatures. In sample 3, the electron and hole mobilities were larger than those in sample 2 and remained largely unchanged with temperature. This may indicate the mobility enhancement brought about by the encapsulating hBN flakes, which eliminate the interface scattering.

\section{Shubnikov de Haas oscillations}

SdH oscillations can be obviously seen at low temperature MR traces (FIG. 2(a) and 2(d)) due to the high carrier mobility and the strong suppression of phonon scattering at low temperatures. The SdH oscillations were extracted from the MR curves by removing the quasi-quadratic backgrounds. The extracted oscillations at different temperatures are shown in FIG. 3(a) and 3(d) for sample 2 and sample 3, respectively. FIG. 3(b) and 3(e) shows the corresponding Fourier transformation (FT) analysis. There were three obvious peaks observed in both sample 2 and sample 3. In sample 2, the three peaks are located at 102 T ($\alpha$), 161 T ($\beta$) and 187 T ($\gamma$). These peak locations almost stay unchanged with temperature. In sample 3, the two peaks located at 84 T and 113 T, and the third peak located at 176 T at 0.02 K and re-located to 162 T above 0.02 K. According to the similarity of locations in sample 2, we assigned a new symbol $\delta$ to the first peak, $\alpha$ to the second peak and $\beta$ to the third peak.

According to the Onsager relation $F=(\Phi_0/2\pi^2)A_F$, where $F$ is the oscillation frequency, $\Phi_0$ is the flux quantum, the cross-sectional area of each Fermi pocket can be obtained: $A_F=0.0097,\ 0.0153,\ 0.0178$ \AA$^{-2}$ for $\alpha$, $\beta$, $\gamma$ pockets in sample 2, respectively. And $A_F=0.008,\ 0.0108,$ $0.0168$ (at 0.02 K) and $0.0154$ (above 0.02 K) \AA$^{-2}$ for $\delta$, $\alpha$ and $\beta$ pockets in sample 3, respectively. It is inaccurate to identify the carrier types of the Fermi pockets based on both the size of each Fermi pocket and the carrier densities obtained from the two-band model fittings, but we can conclude that the electronic structure changes dramatically with thickness, since one pocket ($\gamma$) disappears while a new one ($\delta$) appears in sample 3. In addition, the shift of the $\beta$ pocket with temperature in sample 3 reveals a temperature-sensitive electronic structure of multi-layer WTe$_2$. Besides the Fermi pocket topology, a temperature-induced spin splitting can be explicitly observed in sample 3, as shown in FIG. 3(d). The oscillation peaks double-split with increasing temperature, which is contrary to the common belief that the spin splitting only occurs at low temperature and high magnetic field. What happened here might be due to the breaking of spin-orbit coupling (SOC) with increasing temperature. It has been proved that SOC can be strongly suppressed by temperature in WTe$_2$ \cite{zhang2020crossover}. At low temperature, the SOC is strong so that the applied magnetic field is not sufficient to support an observable Zeeman splitting. At higher temperature, the missing or non-dominant SOC enables the appearance of the Paschen–Back effect at moderate magnetic field \cite{paschen1921liniengruppen, kapitza1938zeeman, abe2019quantitative}.

The Lifshitz-Kosevich (LK) formula is commonly used to analyze the SdH oscillations \cite{shoenberg2009magnetic}. The damping factor shown below in the LK formula is used to describe the temperature dependence of the oscillation amplitude,
\begin{equation}\label{eq:four}
R_{T}=\frac{2\pi^2k_BTm^*/\hbar eB}{\sinh(2\pi^2k_BTm^*/\hbar eB)}
\end{equation}
in which $k_B$ is the Boltzmann constant and $m^*$ is the effective mass. FIG. 3(c) and 3(f) show the fitting results for the Fermi pockets in sample 2 and sample 3, respectively. Since the amplitude of the $\gamma$ peak in sample 2 and $\delta$ peak in sample 3 cannot be reliably extracted above 3.5 K and the $\beta$ peak in sample 3 shifts location, we excluded them from the fitting analysis. The effective mass obtained from fittings are: $m_{\alpha}=0.393\ m_e$ and $m_{\beta}=0.419\ m_e$ for sample 2, and $m_{\alpha}=0.301\ m_e$ for sample 3, where $m_e$ is the free electron mass. We notice that sample 2 shows comparable results with those in bulk, whereas for sample 3, the effective mass is slightly lighter. This might result from the enhanced carrier mobilities due to the encapsulated structure \cite{cai2015drastic, xiang2015multiple, xiang2018thickness}.

\section{Fermi surface anisotropy}

We further measured the angle-dependent MR curves, as shown in FIG. 4(a) and 4(c) for sample 2 and sample 3, respectively. $\theta$ is the angle between the direction of the magnetic field and the normal direction of the sample, as shown in the inset of FIG. 4(b). It is observed that the non-saturating positive MR is strongly suppressed when the magnetic field is parallel to the sample surface. Magnetoresistance oscillation study shows that the peaks in the FT spectrum shift with angle, which indicates the Fermi surface anisotropy between the in-plane and out-of-plane directions. So we further investigated the anisotropy of the Fermi surface through the scaling behavior of MR curves. The scaling behavior \cite{blatter1992isotropic, noto1975simple, thoutam2015temperature} can be expressed as:
\begin{equation}\label{eq:five}
R(H,\theta)=R(\epsilon_{\theta}H)
\end{equation}
with the scaling factor $\epsilon_{\theta}=(\cos^{-2}\theta+\gamma^{-2}\sin^{-2}\theta)^{1/2}$. All the MR curves can be scaled into a single curve as shown in FIG. 4(b) and 4(d), respectively, for both samples. Since the sample resistance is directly related to the effective mass by $R=m^*/ne^2\tau$, where $\tau$ is the relaxation time, the scaling behavior of MR can describe the mass anisotropy $m_{\parallel}/m_{\perp}$, which is the $\gamma$ in the equation. In addition, since $m^*$ manifests the energy band curvature, $\gamma$ alone can also be used to describe the anisotropy of Fermi surface $k_{\parallel}/k_{\perp}$ \cite{noto1975simple, soule1958magnetic, soule1964study, thoutam2015temperature}.

The scaling factors were obtained for both samples at different angles and plotted in FIG. 4(e). By fitting the scaling factor versus angle, we got the anisotropy parameter $\gamma=8.08$ and 2.28 in sample 2 and sample 3, respectively. Interestingly, the anisotropy we got from WTe$_2$ is much smaller than that in graphite \cite{noto1975simple}, a typical layered material. It is even more striking that the anisotropy parameter in the thick sample is larger than the one in the thin sample. That means the electrons in a thinner WTe$_2$ can move more freely along $k_z$ direction than in a thick one. Such a small Fermi surface anisotropy indicates that WTe$_2$ is actually an electronic 3D material. Unlike the isotropic electron gas, however, the 3D electron gas (3DEG) in WTe$_2$ along the stacking direction is strongly modulated by sample thickness. When the thickness reduces to the order of nanometers, a thinner sample may suffer less interlayer scatterings along the stacking direction, thereby making the movement of electrons in this direction less constrained.

\section{Conclusion}

With a thorough magneto-transport study in a series of WTe$_2$ flakes, we have observed a systematic change of electronic structure as a function of the thickness. First, we confirmed that the Kohler’s rule is applicable and responsible for the `turn-on' behavior which normally occurs in Fermi liquid state, thereby ruling out the possibility of a metal-insulator transition. Second, we found that the imbalance of carrier densities took an important role on the suppression of `turn-on' behavior and large positive MR, while the non-saturating characteristic was hardly affected. This might hint at some other origins for such a MR. Third, the SdH oscillation studies further shown the important role of thickness on the Fermi surface in WTe$_2$ and perhaps a temperature-sensitive change in electronic structure. Finally, we reported a thickness-dependent Fermi surface anisotropy, which revealed that WTe$_2$, a typical 2D Van der Waals material, is effectively an electronic 3D material and the anisotropy decreases with decreasing thickness.

\begin{acknowledgments}

This work was supported by UT Dallas SPIRe fund (No. 2108630). Synthesis of WTe$_2$ crystals was supported by DOE BES Award No. DE-SC0014476.

\end{acknowledgments}


%

%


\bibliography{thickness.bib}

\newpage

\begin{figure}
  \centering
  \includegraphics[width=16cm]{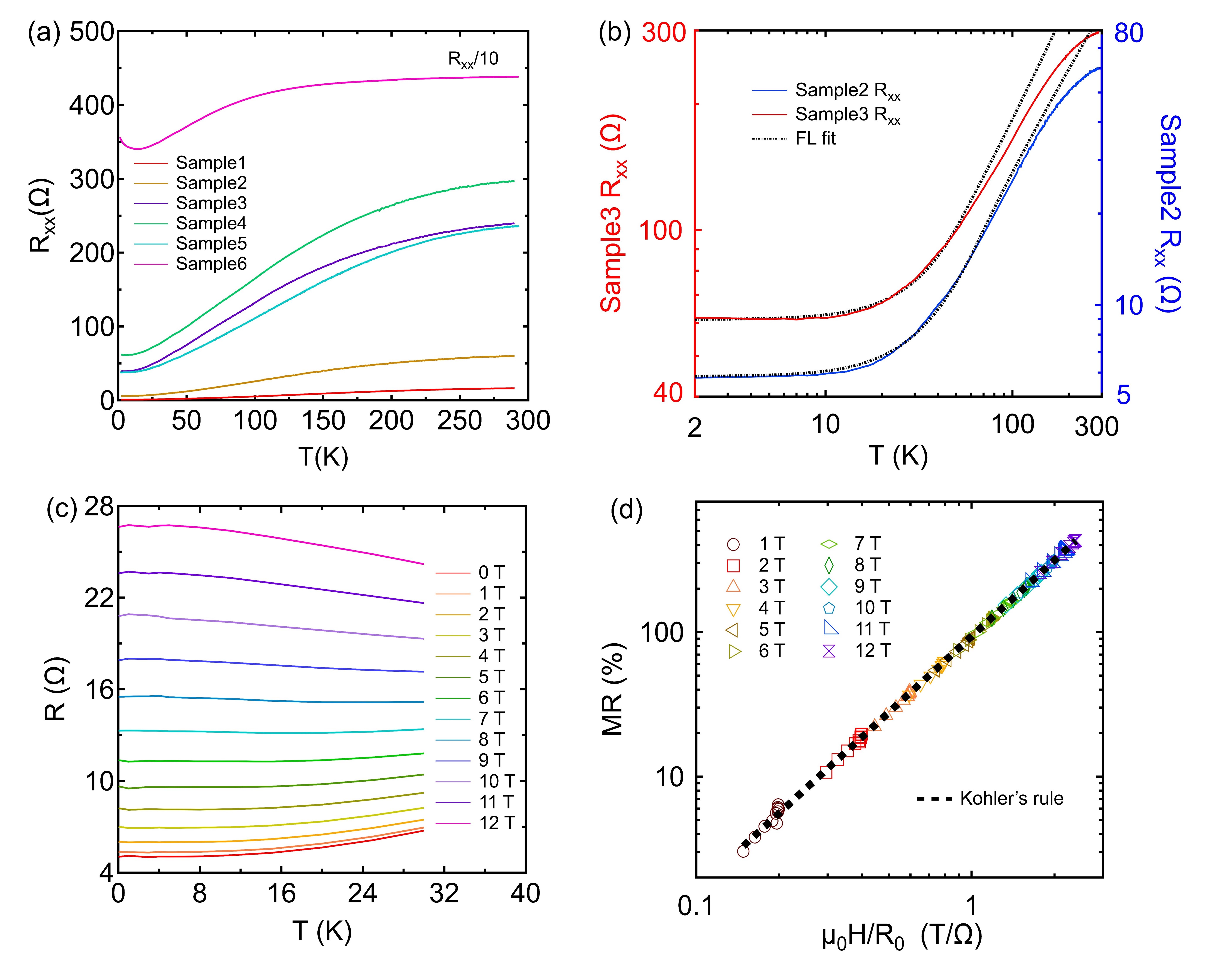}
  \caption{Temperature-dependent resistance measurements. (a) R-T curves of six different devices. Five of them (samples 1-5) show metallic behavior. (b) Fermi liquid fits (dashed lines) for sample 2 and sample 3, respectively, at low temperature regions. (c) R-T curves at various magnetic fields manifest `turn-on' behavior in sample 2. (d) Kohler’s rule scaling of the data in panel (c).}
\end{figure}

\newpage

\begin{figure}
  \centering
  \includegraphics[width=17cm]{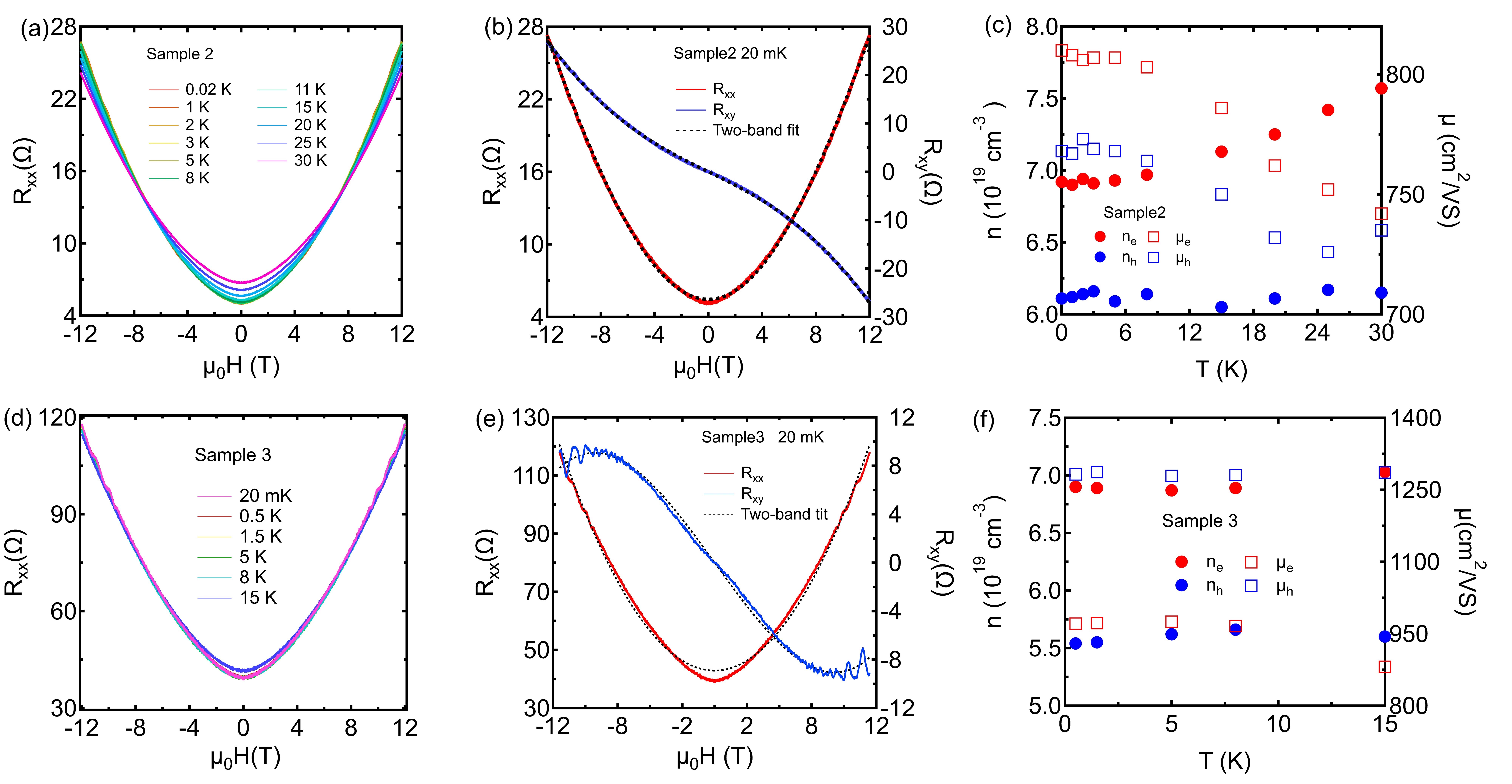}
  \caption{Magnetoresistances and two-band model fittings in samples 2 and 3. (a) MR curves of sample 2 at various temperatures. (b) A representative two-band model fitting result for sample 2 at 20 mK. (c) Carrier densities (solid circles) and mobilities (open squares) for both electrons (red) and holes (blue) extracted from two-band model fittings at different temperatures for sample 2. (d) MR curves of sample 3 at various temperatures. (e) A representative two-band model fitting result for sample 3 at 20 mK. (f) Carrier densities (solid circles) and mobilities (open squares) for both electrons (red) and holes (blue) extracted from two-band model fittings at different temperatures for sample 3.}
\end{figure}

\newpage

\begin{figure}
  \centering
\includegraphics[width=16cm]{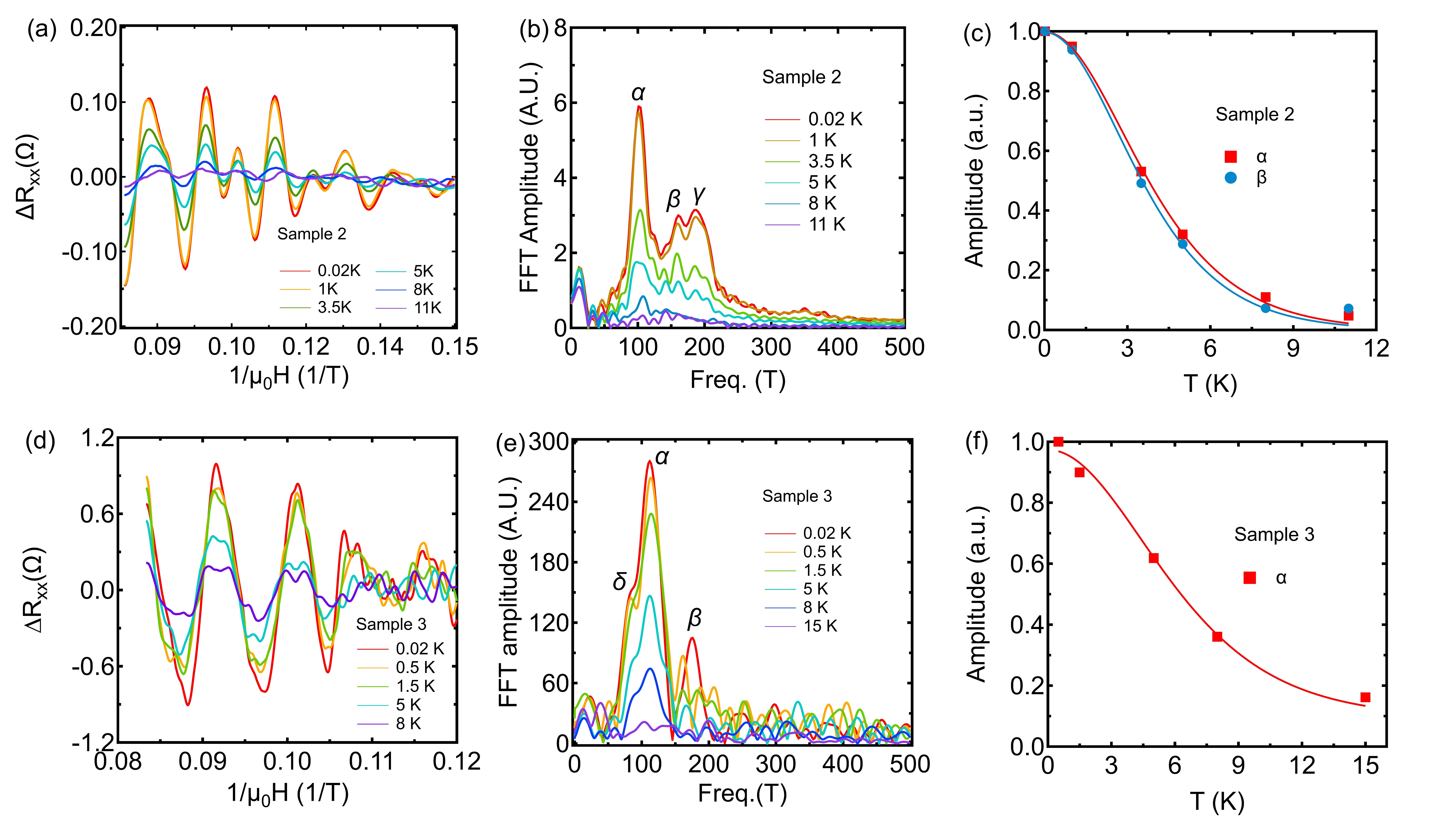}
  \caption{Temperature dependence of SdH oscillations. (a) SdH oscillations extracted from the MR curves for sample 2. (b) FT analysis shows three peaks corresponding to three Fermi pockets in sample 2, which are marked as $\alpha$, $\beta$ and $\gamma$. (c) Temperature dependence of the amplitude of oscillation peaks (square and circle symbols) and the LK fit (solid lines) in sample 2. (d) SdH oscillations extracted from the MR curves for sample 3.(e) FT analysis shows three peaks corresponding to three Fermi pockets in sample 3, which are marked as $\delta$, $\alpha$ and $\beta$. (f) Temperature dependence of the amplitude of oscillation peaks (circle symbols) and the LK fit (solid lines) in sample 3.}
\end{figure}

\newpage

\begin{figure}
  \centering
\includegraphics[width=16cm]{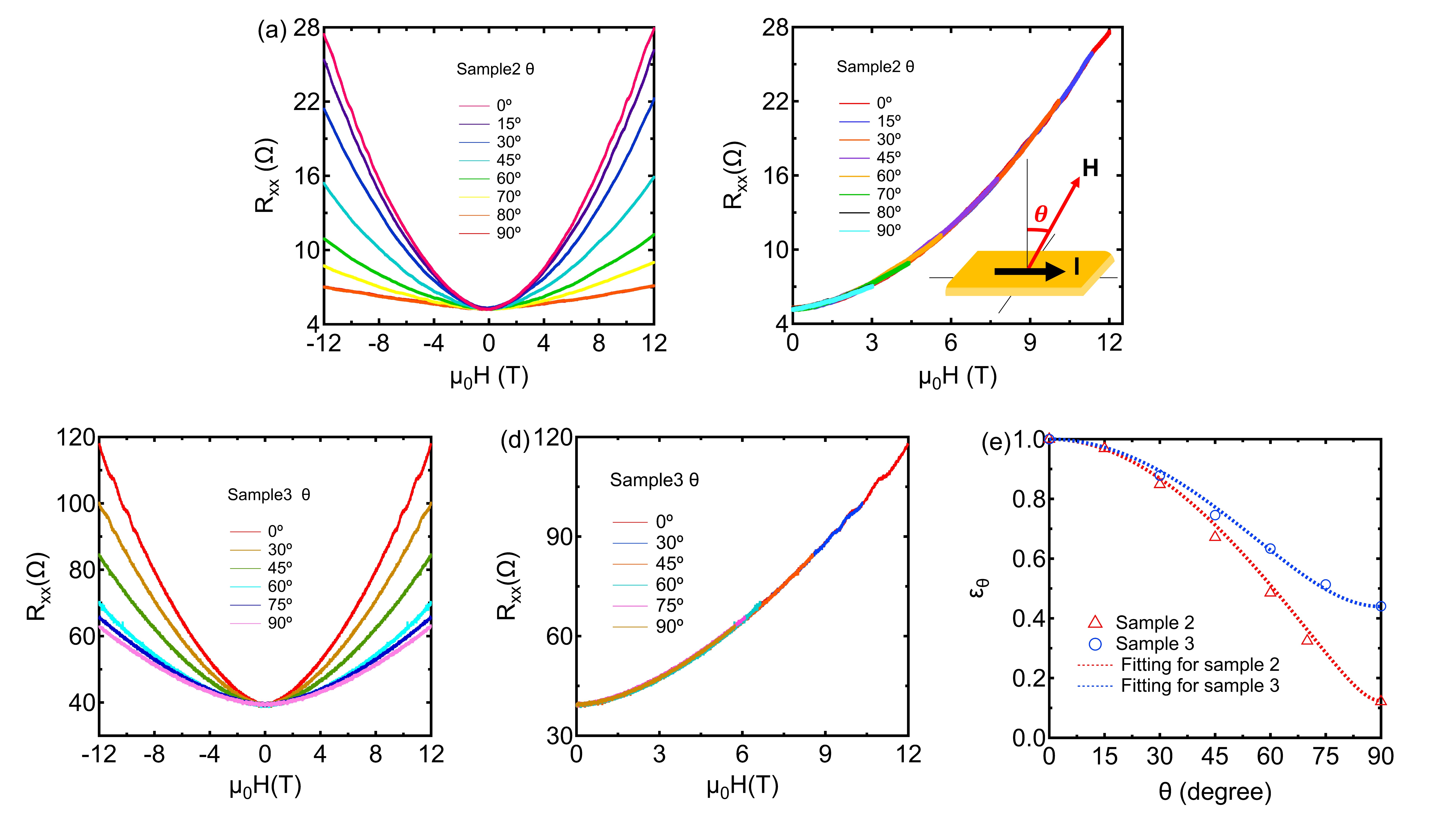}
  \caption{Angle-dependent MR curves and the scaling behaviour. (a) Angle-dependent MR curves in sample 2. (b) The scaling behavior of the MR curves in sample 2. Inset shows the schematic of field orientation. (c) Angle-dependent MR curves in sample 3. (d) The scaling behavior of the MR curves in sample 3. (e) The scaling factors $\epsilon_{\theta}$ at different angles (open symbols) and fittings (dotted lines) for both samples.}
\end{figure}

\end{document}